\documentclass{article}
\usepackage{spconf,amsmath,graphicx}
\usepackage{graphicx}
\usepackage{float} 
\usepackage{subfigure}
\usepackage{array}
\usepackage{multirow}
\usepackage{algorithm}  
\usepackage{algpseudocode}  
\usepackage{amsmath}  
\usepackage[skip=0pt]{caption}
\usepackage{color}
\usepackage{tabularx}
\usepackage[colorinlistoftodos,prependcaption,textsize=tiny]{todonotes}

\title{VSVC: Backdoor attack against Keyword Spotting based on Voiceprint Selection and Voice Conversion}
\name{Hanbo Cai$^1$ \qquad Pengcheng Zhang$^1$ \qquad Hai Dong$^2$ \qquad Yan Xiao$^3$ \qquad Shunhui Ji$^1$ }
\address{$^1$ College of Computer and Information, Hohai University, Nanjing, China\\
$^2$ School of Computing Technologies, RMIT University, Melbourne, Australia\\
$^3$ School of Computing, National University of Singapore, Singapore \\
Email: \{caihanbo, pchzhang, shunhuiji\}@hhu.edu.cn; hai.dong@rmit.edu.au; dcsxan@nus.edu.sg}

\begin{document}
%
\maketitle
\begin{abstract}
Keyword spotting (KWS) based on deep neural networks (DNNs) has achieved massive success in voice control scenarios. However, training of such DNN-based KWS systems often requires significant data and hardware resources. Manufacturers often entrust this process to a third-party platform. This makes the training process uncontrollable, where attackers can implant backdoors in the model by manipulating third-party training data. An effective backdoor attack can force the model to make specified judgments under certain conditions, i.e., triggers. In this paper, we design a backdoor attack scheme based on \underline{V}oiceprint \underline{S}election and \underline{V}oice \underline{C}onversion, abbreviated as \textbf{\textit{VSVC}}. Experimental results demonstrated that VSVC is feasible to achieve an average attack success rate close to 97\%  in four victim models when poisoning less than 1\% of the training data.
\end{abstract}
\section{Introduction}
\label{sec:intro}

Keyword Spotting (KWS) technology based on deep neural networks (DNNs)\cite{KWS} (also known as voice wake-up) has been  applied extensively into the fields of smart home devices (e.g., Apple Home-Pod, Amazon Echo) and smartphone voice assistants (e.g., Siri, Google Assistant, Alexa). By means of KWS, users can easily control various smart devices through specific speech commands. Training a KWS system with satisfactory performance requires high-performing hardware and high-quality labeled speech datasets~\cite{kwsoverview}. Most service providers and individuals outsource hardware platforms and public datasets to third parties, due to the consideration of  cost saving and convenience of training (e.g., cloud computing platforms). However, this will make the providers lose complete control over the training process. Third-party collaborators may bring security risks, where backdoor attacks are one of the threats~\cite{Badnets,zeng2021rethinking,li2022few}.

Backdoor attacks are a new type of DNN model attacking paradigm, which targets their training phase~\cite{backdoorsurvey}. In a backdoor attack, an attacker aims to make a DNN model to only misjudge the data that has certain specific features. By poisoning a small portion of the training data, the attackers induce the DNNs to learn the correlation between backdoor triggers and target labels. In real-world cases, such as autonomous driving~\cite{chen2015deepdriving}, smart healthcare~\cite{smarthealth}, etc, an attacker who knows the backdoor can implicitly instruct the target device to operate in a specific way. This would generate substantial security risks in some use scenarios.

So far a significant mount of research efforts have been devoted to backdoors of image and text classification~\cite{Badnets,li2021backdoor,InvisibleBackdoor,Learnabletextualbackdoor,MSTBackdoor,liu2022piccolo}. In contrast, few attention has been paid to backdoor attacks against speech recognition. Liu et al.~\cite{TANN} reversed a neural network to generate triggers and poisoned data, and then retrained the speech recognition model to implant a backdoor. Zhai et al.~\cite{Zhai} devised a backdoor attack against speaker verification systems. Based on a clustering-based approach, they planted a backdoor in a speaker verification system. Stefanos et al.~\cite{canyouhear} designed a highly stealthy backdoor attack scheme based on ultrasonic pulses. Qiang et al.~\cite{qiang2022opportunistic} designed a passive and opportunistic trigger using ambient noise, such as music and common noise in life, as a trigger for backdoor attacks. Shi et al.~\cite{shi2022audio} developed an optimization scheme to generate position-independent, unnoticeable, and robust audio triggers. Ye et al.~\cite{Ye2019adversarial} designed an audio trigger embedded with hidden information using audio steganography. 

However, these methods still suffer from the following limitations:

\begin{itemize}
\item \textbf{\textit{Most methods ignore the invisibility of a trigger to human ears.}} The triggers generated by most methods are a piece of human-incomprehensible noise. Human ears can distinguish between benign and poisoned samples injected with triggers, making such attacks highly detectable~\cite{canyouhear}.

\item \textbf{\textit{Existing inaudible triggers have significant limitations.}} The existing inaudible triggers mainly utilize the insensitivity of human ears to achieve stealthiness. However, such methods often contain limitations during implementation. According to the Nyquist sampling theorem~\cite{nyquist1}, if an ultrasonic pulse trigger (the signal frequency is 21kHz in \cite{canyouhear}) is to be recognized without distortion, the sampling rate of the speech model should be at least two times greater than the highest frequency in the signal (the sampling rate is 44.1kHz in \cite{canyouhear}); otherwise, such attacks will be ineffective. In addition, a low-pass filter can easily filter out inaudible ultrasonic pulse triggers. Ultrasonic pulse triggers are also challenging when attacking LSTMs~\cite{canyouhear}. Employing ambient sounds as triggers is also a puzzle. The challenge is mainly reflected in how to deduce a suitable location for trigger implantation and choose an appropriate ambient sound for triggering~\cite{shi2022audio}.

\item \textbf{\textit{Most existing research ignores attacking multiple target classes.}} Existing methods ~\cite{TANN,canyouhear,qiang2022opportunistic,shi2022audio} only focus on attacking single target classes and are only triggered by single backdoors, i.e., a one-to-one attack. Under multi-target attacks, attackers can implant multiple backdoors through different triggers, making the attack scenarios more extensive. However, in speech recognition, the triggering paradigm for backdoor attacks against multiple target classes has yet to be studied.

\end{itemize}

We propose a backdoor attack scheme based on voice conversion to realize stealthy multi-target attacks. We adopt voice conversion to transform the timbre of a speech. The timbre can be used as a special trigger. Such a trigger is stealthy to human ears, which can address the aforementioned limitation of the existing inaudible triggers. The main contributions of this work are as follows:

\begin{itemize}
\item We designed a trigger paradigm aiming at converting the timbre of speech voice.
  Extensive experiments demonstrated that our method is effective and stealthy in attacking KWS at low poison rates ($\leq 1\%$). Our triggers are more stealthy than ~\cite{TANN,qiang2022opportunistic,shi2022audio,Ye2019adversarial} since no incomprehensible noises are produced.

\item The trigger paradigm based on voice conversion can change the timbre features of the entire speech, so it is not limited by the sampling rate. On LSTM, our method can achieve a 92\% attack success rate when poisoning only 50 samples, which is 72\% higher than the baseline method.

\item Since timbre, as a critical acoustic feature, has rich forms of expression, we can implant different timbre triggers for various attack targets to achieve multi-target backdoor attacks. In our experiment, we demonstrate implanting three triggers to launch attacks on three target labels, i.e., many-to-many attacks.

\end{itemize}

\section{Preliminaries}
\noindent \textbf{Voiceprint extraction.} Voiceprint extraction is to select the only stable and reliable features from the voice that express the identity of a speaker~\cite{voiceprint1962}. The selected features can effectively distinguish different speakers, and the voiceprint features of each speaker are unique and stable. Experiments show that the more significant difference in timbre, the smaller the similarity of voiceprints between two speakers. Due to the excellent performance of the x-vector~\cite{xvector}, this paper chooses the x-vector to extract the voiceprint features from the speech dataset.

\noindent \textbf{Voice Conversion.}  Voice conversion~\cite{voiceconversion} refers to keeping the original semantic features of voice unchanged while varying its personalized features. A speech contains a wealth of features, the most important of which are semantic features and voice personalization features (i.e., timbre, accent, speaking style, etc.). This paper mainly uses the StarGANv2-VC~\cite{li2021starganv2} framework to convert timbre. StarGANv2-VC performs voice conversion using different timbre classes as the conditional input for the generator and discriminator, thus realizing the conversion from one-to-one to $N$-to-$N$.

\noindent \textbf{Threat Model.} We assume that an attacker knows nothing about the structure of a DNN model. Nonetheless, the attacker has full access and modification rights to the training dataset to craft the poisoned data. Such scenarios are common in  training platform or training data outsourcing~\cite{commonvoice,TSAA}.

\noindent \textbf{Attacker's Goals.} The attacker has two main goals~\cite{backdoorsurvey}. The first goal is to improve the triggering efficiency of the backdoor. The higher the triggering efficiency, the more likely the attack will succeed, and the fewer queries need to be initiated. In this regard, some application scenarios may limit the number of queries, thus making the attack ineffective. The second goal is stealth. The poisoned model should behave like a normal clean model when being fed with samples that do not contain triggers. The triggers should be as indistinguishable for human ears as possible.

\begin{figure}[htbp]
\centering
\includegraphics[width=\columnwidth]{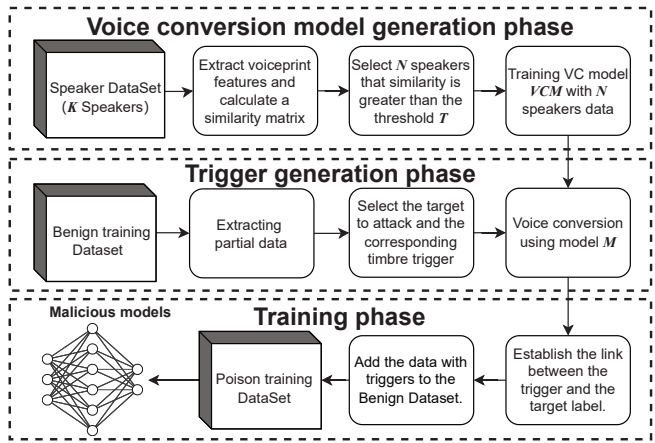}
\vspace{-3mm}
\caption{The overall framework of the method}
\vspace{-6mm}
\label{fig:process}
\end{figure}

\section{The proposed approach}
\vspace{-3mm}
\label{sec:approach}
Spectrogram can be used to represent the spatial and temporal features of a speech. Existing state-of-the-art speech recognition models usually use spectrograms as input features~\cite{deepspeech,deepspeech2}. The traditional trigger paradigm rarely considers the temporal features, which usually targets single speakers' speeches. It is difficult to attack effectively and carry out multi-target attacks.

To overcome the aforementioned shortcomings of the previous studies, we propose a voice-conversion-based triggering paradigm. The framework of voice conversion is StarGANv2-VC~\cite{li2021starganv2}. Only the timbre features of audio are varied in its converted version, with the resultant triggers stealthy to human ears.

As shown in Figure \ref{fig:process}, the framework comprises three phases - a voice conversion model generation phase (see Algorithm \ref{alg:vc_gen}), a trigger generation phase, and a poisoning training phase.
\vspace{-3mm}
\subsection{Voice conversion model generation phase}
\hspace{1.5em}\textbf{(1) Speaker corpus voiceprint similarity calculation.} The first step is to extract the timbre features of $N$ speakers from high-quality speaker corpus as triggers. We choose the \textit{VCTK}~\cite{VCTK} as the timbre corpus to train the voice conversion model. For multi-target attacks, the triggers should be as different as possible; otherwise the effectiveness of the attack will be weakened. Therefore, the extracted speaker timbre features should also have remarkable distinction. The corpus contains 109 speakers, each with a unqiue timbre style. We use an x-vector to extract the voiceprint feature vector of each speaker (Lines 2-4 of Algorithm \ref{alg:vc_gen}) and then generate a voiceprint similarity matrix by Euclidean distance (Lines 5-9 in Algorithm \ref{alg:vc_gen}).

\textbf{(2) Voice conversion model training.} According to the voiceprint similarity matrix, if we aim to attack $N$ targets, we select the timbres of $N$ speakers ($S_1$,...,$S_N$) as triggers. To ensure sufficient discrimination, the voiceprint similarity between the $N$ speakers should be greater than the threshold $T$ (Line 10 to 14 in Algorithm \ref{alg:vc_gen}). The voice conversion model is then trained using the StarGANv2-VC framework on the $N$ speaker corpus (Line 15 in Algorithm \ref{alg:vc_gen}). For single-target attacks, we randomly select one speaker's data for training (Lines 17-18 of Algorithm \ref{alg:vc_gen}). Finally, we obtain the trained voice conversion model $VCM$.

\begin{algorithm}[ht]
\caption{Voice conversion model generation algorithm}
\label{alg:vc_gen}
  \begin{algorithmic}[1]
    \Require Corpus Dataset with $K$ Speakers: $D_K$; \par
      \hspace{0.4em} Number of attack targets: $N$; \par
      \hspace{0.4em} Threshold of smilarity $T$;
    \Ensure  Voice conversion Model: $VCM$;
    \If {$N$ \textgreater 1}
        \Statex \textbf{\textit{\qquad //Extract the voiceprint features of each speaker.}}
        \For{$i$ in range($K$)}
        \State   $Spk\_embed_i$ ← X-vector($Speaker_i$);
        \EndFor
        \Statex \textbf{\textit{\qquad //Generate a similarity matrix using Euclidean
        \Statex \hspace{2.2em} distance.}}
        \For{$i$ in range($K$)}
            \For{$j$ in range($K$)}
            \State $SM[i][j]$ = EucDist($Spk\_embed_i$,$Spk\_embed_j$)
            \EndFor
        \EndFor
        \Statex \textbf{\textit{\qquad //Select N candidate speakers' data for training the 
        \Statex \hspace{2.2em} voice conversion model.}}
        \For{$i$ in range($K$)}
            \If { $len(Cand\_speakers)$ \textless $N$ \textbf{AND} \par  \hspace{0.5em}$SM$[$Spk\_embed_i$][$Cand\_speakers$] \textgreater $T$}
                \State $Cand\_speakers.append(Speaker\_i)$
            \EndIf
        \EndFor
        \State $VCM$ = Train($StarGANv2$-$VC$,$Cand\_speakers$)
    \Else
        \State $Speaker$ = Random\_Select($D_K$)
        \State $VCM$ = Train($StarGANv2$-$VC$,$Speaker$)
    \EndIf
    \State \textbf{return} $VCM$;
  \end{algorithmic}
\end{algorithm}
\vspace{-3mm}
\subsection{Trigger generation phase}
\hspace{1.5em}\textbf{(1) Timbre Trigger pairing.} Given the target label $i$ to be attacked, a speaker $S_i$ is selected from the $N$ speakers, and the speaker $S_i$' timbre feature is used as the trigger for the target label $i$.

\textbf{(2) Timbre trigger injection.} We employ the voice conversion model $VCM$ to convert the benign audio data into the timbre of the corresponding speaker $S_i$. This benign audio data is implanted with timbre triggers.

In summary, the first step in the generation phase is to construct a voiceprint similarity matrix for the speakers in the corpus. Then we select $N$ speakers whose voiceprint similarity with each other is greater than the threshold $T$. Finally, we use the StarGANv2-VC framework to learn the timbre features of these $N$ speakers. Different timbres features can generate different triggers to achieve multiple backdoor attacks.
\vspace{-8mm}
\subsection{Training phase}
\hspace{1.5em}\textbf{(1) Dataset poisoning.} Given the poisoning rate parameter \textit{\textbf{p}},  the corresponding timbre trigger is injected into each target to be attacked. It will generate \textbf{\textit{p\% }}poisoning data and constructs the poisoning training set. \textbf{\textit{p}} is an essential hyperparameter in the attacking process, and the value of \textbf{\textit{p }}has a crucial impact on the attacking results. There is also a trade-off between attacking performance and stealthiness.

\textbf{(2) Poison Model Training.} Following the above steps, \textbf{\textit{N}} triggers have been associated with \textbf{\textit{N}} target labels, forming a poisoning training dataset. It will be unitized to train a malicious model.
\vspace{-3mm}
\section{Experimental results}
\vspace{-3mm}
\label{sec:typestyle}
\subsection{Experimental Setting}
\vspace{-3mm}
\textbf{Dataset Description.} We use the Google Speech Command dataset~\cite{speechcmd} as our experimental dataset. The dataset contains 65,000 audios, each of which is labelled by a single word (30 words in total). Each word file is a one-second speech clip with a 16kHz sampling rate We select 23,682 audios with 10 labels (``yes", ``no", ``up ", ``down", ``left", ``right", ``on ", ``off", ``stop", and ``go") for our experiments. We set the ratio of the training set to the test set to 9:1.

\noindent\textbf{Victim Models.} Our experiments are conducted on four KWS classification networks: LSTM~\cite{LSTM}, ResNet18~\cite{resnet}, VGG16~\cite{VGG} and WideResNet-10-10~\cite{wideresnet}, all of which have excellent classification performance in speech command recognition challenges~\cite{srchallenge}. We slightly modify their network input structure to adapt to the spectrogram input requirements.

\noindent\textbf{Training Setup.} We extract the log-Mel spectrogram of each audio sample as an input feature, which can graphically characterize a person's speech feature in a combination of temporal and spatial dimensions. Therefore, we can also use the image classification networks to recognize speech data. We use cross-entropy~\cite{crossentropy} as the loss function and SGD~\cite{SGD} as the optimizer. The learning rate of LSTM is set to 0.005, and the learning rates for the rest of the models are set to 0.01. For our attack, we evaluate the performance trend of the poisoning rate \textbf{\textit{p}} ranging from 0.2\% to 2\%.

\noindent\textbf{Baseline Selection.} We select the ultrasonic pulse method in ~\cite{canyouhear} as the baseline for comparison. According to the findings in ~\cite{canyouhear}, short ultrasonic pulses are sensitive to implant locations and have a lower Attack Success Rate (ASR). Therefore, to balance the stealthiness and ASR, we set the duration of the long ultrasonic pulse to 100 ms. The sampling rate of the baseline method must be at least 44.1kHZ; otherwise the attack will be ineffective.

\noindent\textbf{Evaluation metrics.} We evaluate our approach in terms of Attack Success Rate (ASR), Accuracy Variance (AV), and human verification. To assess the ASR of an attack target, we select a part of the data from the test set, which has different labels from the attack target label. Then we inject our pre-designed timbre trigger into this part of the data (that is, performing voice conversion on this part of data), and calculate the percentage of successful backdoors triggered by these data when feeding into the model. AV refers to the model's prediction accuracy variance for benign samples before and after the backdoor attack. The higher the ASR and the smaller the AV, the better the attack performance. Human verification assesses whether the backdoor audio is natural and intelligible.

\noindent\textbf{Evaluation Setup.} We select one (single-target attack) or $N$ (multi-target attack) target labels to attack for each model. All the experiments are repeated five times to reduce the effect of randomness. We calculate the ASR against each target in the multi-target attack and averaged it. In addition, for multi-target attacks, we compare the results of randomly selecte timbres and selected timbres based on the voiceprint similarity (\textbf{VSVC}) for ablation studies. In the experiment, we set the value of $N$ to 3.

\begin{figure*}[htbp]
\subfigure[]{
\label{fig:ASR(a)}
\centering
\begin{minipage}[t]{0.25\textwidth}
\hspace{-0.2cm}\includegraphics[width=4.6cm]{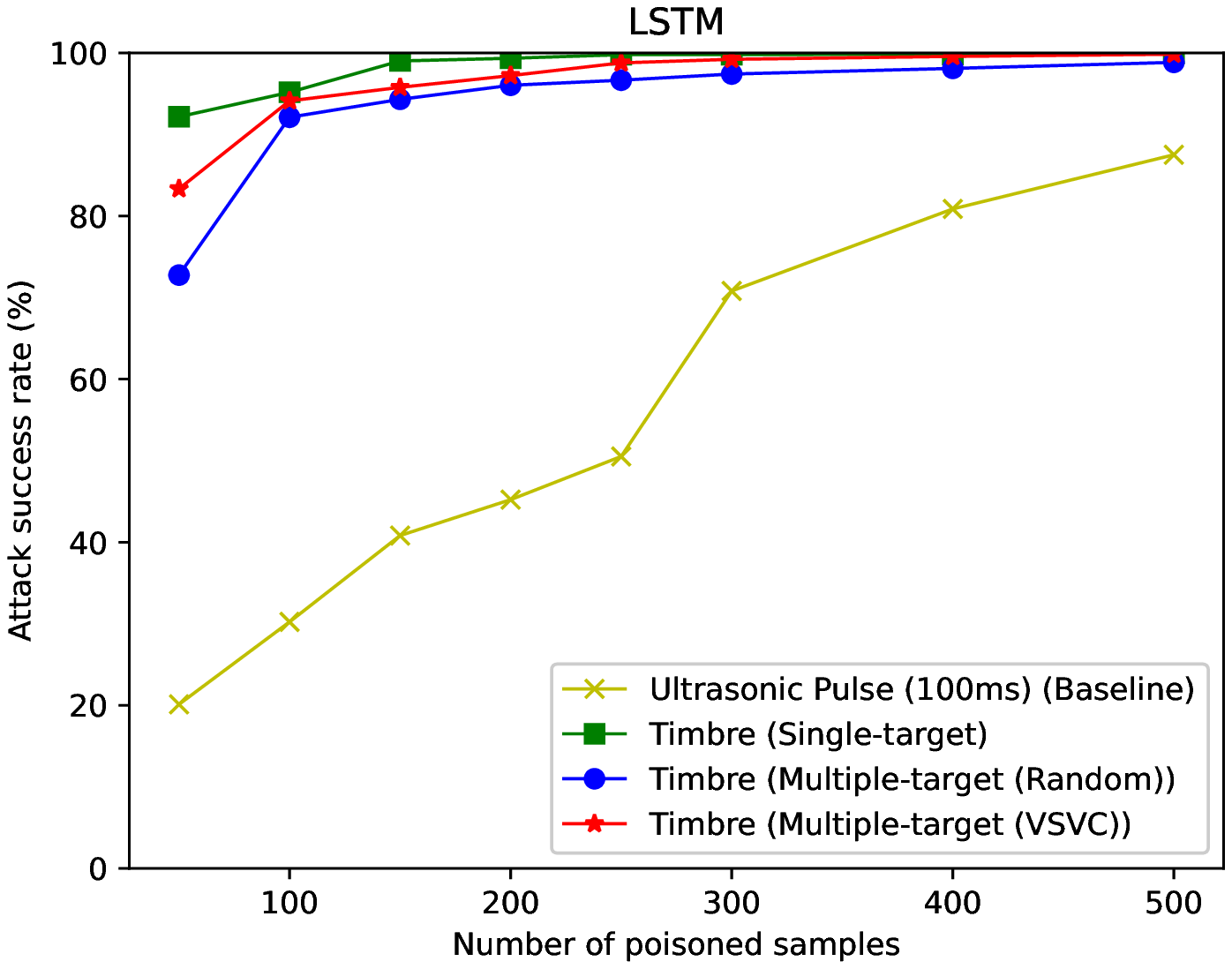}
\end{minipage}%
}%
\subfigure[]{
\label{fig:ASR(b)}
\begin{minipage}[t]{0.25\textwidth}
\hspace{-0.2cm}\includegraphics[width=4.6cm]{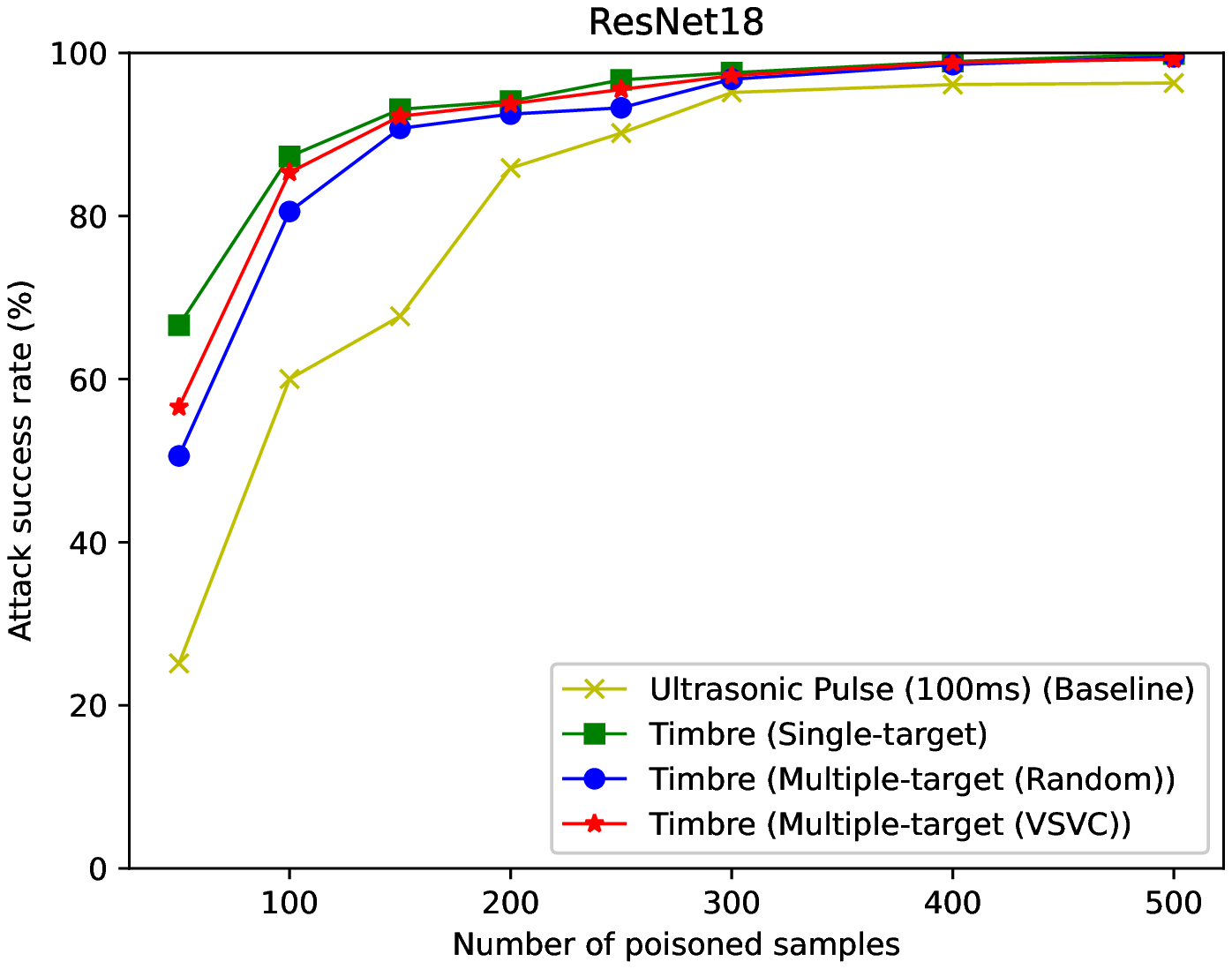}
\end{minipage}%
}%
\subfigure[]{
\label{fig:ASR(c)}
\begin{minipage}[t]{0.25\textwidth}
\hspace{-0.2cm}\includegraphics[width=4.6cm]{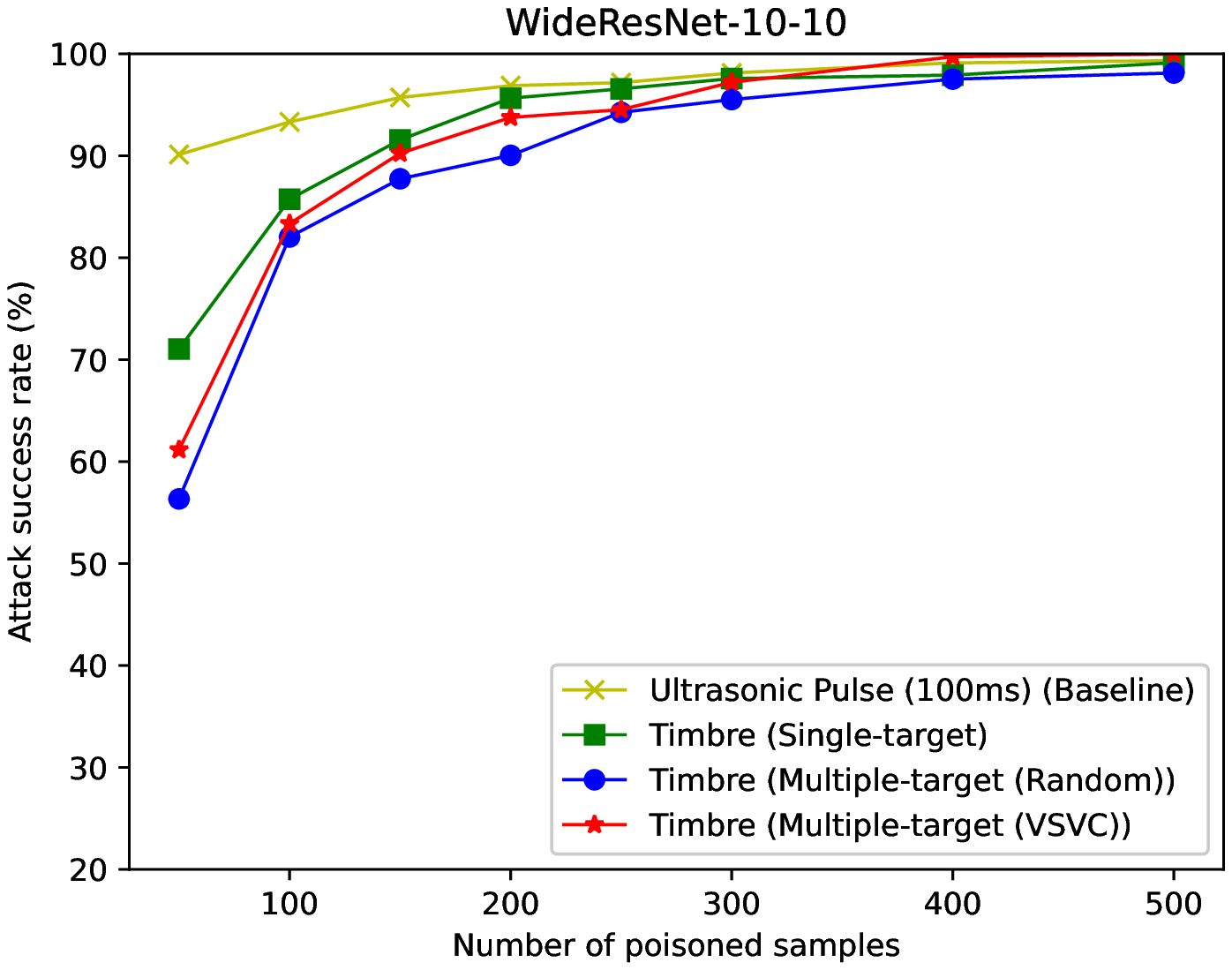}
\end{minipage}
}%
\subfigure[]{
\label{fig:ASR(d)}
\begin{minipage}[t]{0.25\textwidth}
\hspace{-0.2cm}\includegraphics[width=4.6cm]{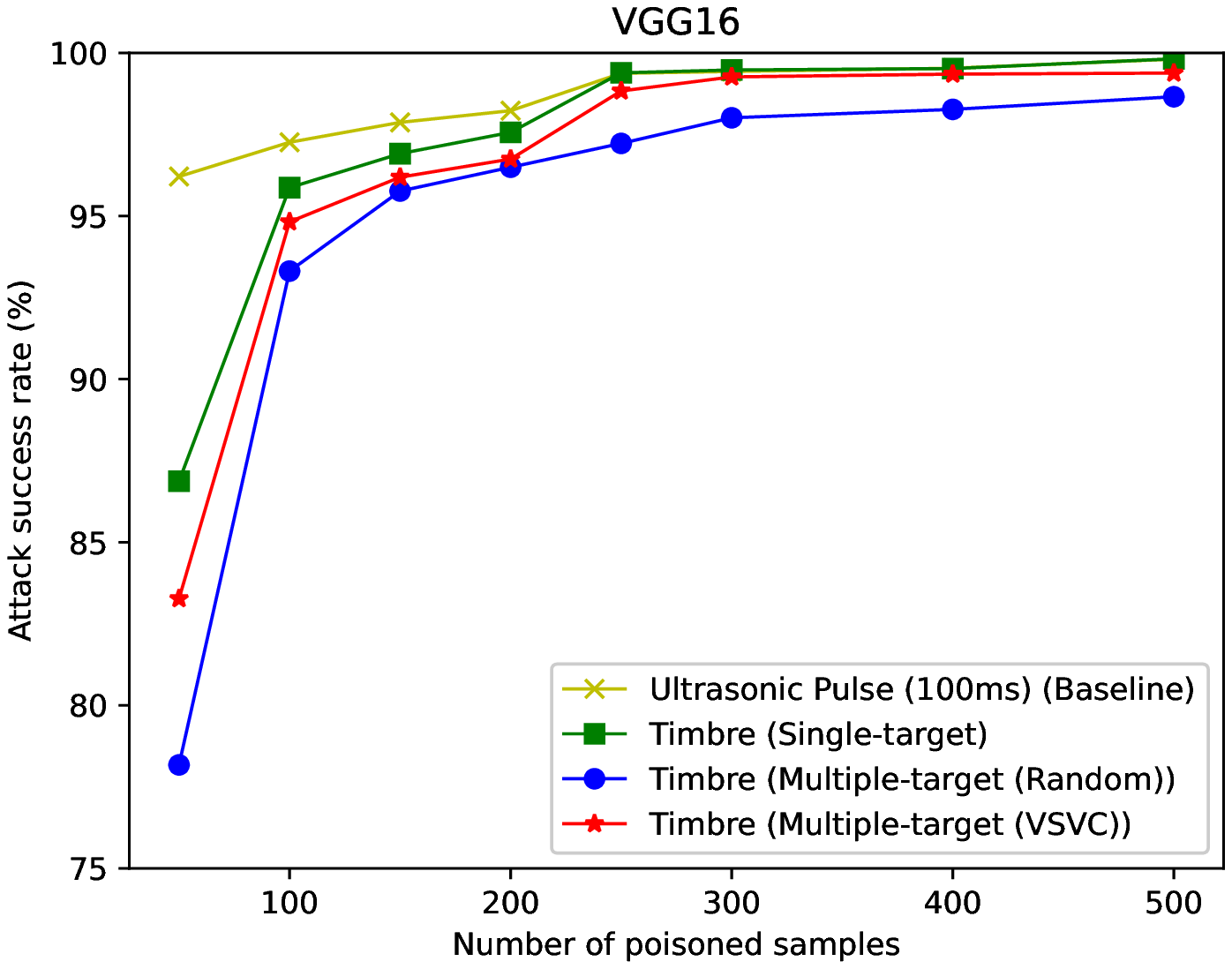}
\end{minipage}
}%
\vspace{-1mm}
\caption{Attack Success Rate on KWS (\%)}
\label{fig:ASR}
\vspace{-5mm}
\end{figure*}

\begin{table*}[htbp]
\caption{Comparison of accuracy variance before and after backdoor attacks (\%).}
\centering
\label{tab:acccompare}
\newcolumntype{?}{!{\vrule width 1.5pt}}
\tabcolsep=0.29cm
\begin{tabular}{p{1.5cm}<{\centering}c|cccccc}
\hline
\multirow{2}{*}{model} & \multirow{2}{*}{original accuracy} & \multicolumn{6}{c}{Accuracy Variance for different numbers of poisoned samples (VSVC / Baseline)}                                                        \\ \cline{3-8} 
                       &                                    & \multicolumn{1}{c}{50} & \multicolumn{1}{c}{100} & \multicolumn{1}{c}{200} & \multicolumn{1}{c}{300} & \multicolumn{1}{c}{400} & \multicolumn{1}{c}{500} \\ \hline
LSTM                   & 92.75                              & 0.08 / 0.32            & 0.11 / 0.47             & 0.35 / 0.54             & 0.50 / 0.59             & 0.74 / 0.61             & 0.78 / 0.70             \\
VGG16                  & 95.99                              & 0.03 / 0.05            & 0.46 / 0.37             & 0.34 / 0.63             & 0.35 / 0.54             & 0.58 / 0.65             & 0.66 / 0.75             \\
ResNet18               & 94.74                              & 0.24 / 0.51            & 0.39 / 0.27             & 0.51 / 0.56             & 0.43 / 0.54             & 0.78 / 0.71             & 0.62 / 0.81             \\
\makebox[0.12\textwidth][c]{WideResNet-10-10}             & 93.81                              & 0.29 / 0.34            & 0.23 / 0.32             & 0.47 / 0.53             & 0.54 / 0.52             & 0.66 / 0.69             & 0.86 / 0.77             \\ \hline
\end{tabular}
\vspace{-6mm}
\end{table*}
\vspace{-4mm}
\subsection{Results and analyses}
\vspace{-2mm}
As shown in Figure \ref{fig:ASR} and Table \ref{tab:acccompare}, timbre as a trigger can successfully attack all victim models. In particular, excellent attack performance has been obtained on LSTM. Although, the ASR of our method  on VGG16 and WideResNet-10-10 is lower than the baseline method at a low poisoning rate. However, as mentioned in the introduction, the baseline method has stringent requirements on the sampling rate (it must be greater than 44.1kHZ) and is extremely easy to be filtered. In comparison, our method is not subject to such limitations. In addition, the performance of this method is consistent, i.e., the ASR quickly converges to a high value (\textgreater95\%) with the increased number of poisoned samples, while the AV is maintained in a small range.
\vspace{-3mm}
\subsubsection{Single Target Attack Success Rate Analysis}
\vspace{-2mm}
As shown in Figure \ref{fig:ASR}, the timbre trigger can achieve the best performance under a single target attack. Especially on LSTM, when the poisoning rate reaches 0.63\% (150 samples in Figure \ref{fig:ASR(a)}), ASR can reach to approximately 100\%, which is 58.2\% higher than the baseline method. This addresses the shortage problem of temporal baseline models. Although the experimental data used in this paper are all short speech clips with a duration of one second, after feature extraction, a spectrogram with a length of 87 frames can be obtained, which is essentially a temporal sequence. The timbre is rich in features, which are coherent in the time dimension. Thus, the change of timbre can significantly affect the classification results of LSTM. The baseline method only considers the characteristics of spatial dimension  but ignores the features of  time dimension. Therefore, it cannot effectively attack LSTM.

For ResNet18, under single-target attacks, our method can outperform the baseline method on ASR by 38.8\% at a low poisoning rate, reflecting more robust performance. The following reasons may cause this. First, ResNet strictly adheres to the processing of extracting global features from local features during training, in addition to performing global averaging pooling and batch normalization to reduce the local features in the higher levels significantly. Second, the shortcut connection of ResNet18 leads to only a small number of features useful to residual blocks learning. Given that most residual blocks can only provide a small amount of information, some local features will be lost. Third, since the gradient can directly pass through the shortcut instead of being forced to pass through the residual block during backpropagation, this will cause only a very limited number of layers to learn useful knowledge. Hence, most of the layers contribute little to the final result, which causes many low-level features to be washed out. The baseline method cannot achieve satisfactory ASR on ResNet18, because it only modifies local features. Since our method greatly compensates for the above problems by changing the global features,  it can successfully attack ResNet18 at a meager poisoning rate.

The experimental results on WideResNet-10-10 further prove this view. WideResNet-10-10 improves the capture of local features by increasing the width and reducing the depth, which allows the residual blocks to learn as much useful local information as possible.  Therefore, on the WideResNet-10-10 network, the attacking performance of all the methods is improved.

Finally, as a typical deep convolutional neural network, we conduct experiments on VGG16. VGG16 is better at extracting local features and considering spatial feature associations, while the implantation of timbre triggers mainly affects sequential features. As a result, the attacking performance of our method on VGG16 is slightly lower than that of LSTM.

Since the baseline method mainly affects local features, it has more robust representation capabilities for models, such as WideResNet-10-10 and VGG16, that are more effective at extracting local features. Therefore, the baseline method has a higher ASR than ours at a low poisoning rate.
\vspace{-4mm}
\subsubsection{Multiple Target Attack Success Rate Analysis}
\vspace{-2mm}
We  also conduct experiments to verify our method's feasibility for multi-target attacks. Under the multi-target attack, the ASR is slightly lower than that under the single-target attack (see Figure 2). This is because when multiple timbre triggers are implanted, if the timbre feature of the triggers is similar, it may cause unavoidable interference. This leads to a decrease in the ASR. At a low poisoning rate, this phenomenon will be more obvious. When the poisoning rate is low, the model cannot sufficiently learn the feature differences between the two timbres. Such problems can be alleviated when the poisoning rate increases.

To further verify the above point, we compared the experimental results of random selection and selection of multiple timbre triggers based on the voiceprint similarity (\textbf{VSVC}). The comparison of the two results shows that, in the multi-target attack, generating multiple timbre triggers based on voiceprint similarity selection leads to better attacking performance than random selection, especially at low poisoning rates. This is because VSVC can avoid two similar timbres from attacking different targets, resulting in more distinctive features for each trigger, which makes multi-target attacks more robust. 
\vspace{-3mm}
\subsubsection{Accuracy Variance Analysis}
\vspace{-2mm}
In addition, we evaluate the prediction accuracy of the victim models on benign samples before and after the backdoor attacks. The average accuracy variance of VSVC is 0.45\%, which is lower than that of the baseline (0.53\%). This can be viewed as insignificant.
\vspace{-3mm}
\subsubsection{Human Validation}
\vspace{-2mm}
We recruit 20 tertiary students who pass College English Test (CET) 4 as volunteers to conduct human verification. We randomly played 20 audios for each volunteer, including 10 backdoor and 10 benign audios, and asked the volunteers to distinguish between these two types of audio. The results showed that 100\% of the volunteers could not distinguish between the two types of audio. Combined with the Accuracy Variance result, our attack is approved to be stealthy.
\vspace{-3mm}
\section{Conclusions}
\vspace{-2mm}
This paper explores how to conduct a backdoor attack against keyword spotting. To address the limitations of existing backdoor attack methods against speech recognition, we propose a backdoor attack scheme combining voiceprint selection and voice conversion. This method is not constrained by the sample rate, which has better stealthiness and can realize multi-target attacks. Moreover, our method can effectively attack temporal layers (such as LSTM) compared to the baseline method.

\vfill\pagebreak

\bibliographystyle{IEEEbib}
\bibliography{mybib}

\begin{thebibliography}{10}

\bibitem{KWS}
Tara~N. Sainath and Carolina Parada,
\newblock ``Convolutional neural networks for small-footprint keyword
  spotting,''
\newblock in {\em {INTERSPEECH}}. 2015, pp. 1478--1482, {ISCA}.

\bibitem{kwsoverview}
Iv{\'a}n L{\'o}pez-Espejo, Zheng-Hua Tan, John Hansen, and Jesper Jensen,
\newblock ``Deep spoken keyword spotting: An overview,''
\newblock {\em IEEE Access}, vol. 10, pp. 4169--4199, 2022.

\bibitem{Badnets}
Tianyu Gu, Kang Liu, Brendan Dolan{-}Gavitt, and Siddharth Garg,
\newblock ``Badnets: Evaluating backdooring attacks on deep neural networks,''
\newblock {\em {IEEE} Access}, vol. 7, pp. 47230--47244, 2019.

\bibitem{zeng2021rethinking}
Yi~Zeng, Won Park, Z~Morley Mao, and Ruoxi Jia,
\newblock ``Rethinking the backdoor attacks' triggers: A frequency
  perspective,''
\newblock in {\em Proceedings of the IEEE/CVF International Conference on
  Computer Vision}, 2021, pp. 16473--16481.

\bibitem{li2022few}
Yiming Li, Haoxiang Zhong, Xingjun Ma, Yong Jiang, and Shu-Tao Xia,
\newblock ``Few-shot backdoor attacks on visual object tracking,''
\newblock in {\em International Conference on Learning Representations}, 2022.

\bibitem{backdoorsurvey}
Yiming Li, Yong Jiang, Zhifeng Li, and Shu-Tao Xia,
\newblock ``Backdoor learning: A survey,''
\newblock {\em IEEE Transactions on Neural Networks and Learning Systems}, pp.
  1--18, 2022.

\bibitem{chen2015deepdriving}
Chenyi Chen, Ari Seff, Alain Kornhauser, and Jianxiong Xiao,
\newblock ``Deepdriving: Learning affordance for direct perception in
  autonomous driving,''
\newblock in {\em Proceedings of the IEEE international conference on computer
  vision}, 2015, pp. 2722--2730.

\bibitem{smarthealth}
Stephanie~B Baker, Wei Xiang, and Ian Atkinson,
\newblock ``Internet of things for smart healthcare: Technologies, challenges,
  and opportunities,''
\newblock {\em Ieee Access}, vol. 5, pp. 26521--26544, 2017.

\bibitem{li2021backdoor}
Yiming Li, Tongqing Zhai, Yong Jiang, Zhifeng Li, and Shu-Tao Xia,
\newblock ``Backdoor attack in the physical world,''
\newblock {\em arXiv preprint arXiv:2104.02361}, 2021.

\bibitem{InvisibleBackdoor}
Yuezun Li, Yiming Li, Baoyuan Wu, Longkang Li, Ran He, and Siwei Lyu,
\newblock ``Invisible backdoor attack with sample-specific triggers,''
\newblock in {\em ICCV}. 2021, pp. 16443--16452, {IEEE}.

\bibitem{Learnabletextualbackdoor}
Fanchao Qi, Yuan Yao, Sophia Xu, Zhiyuan Liu, and Maosong Sun,
\newblock ``Turn the combination lock: Learnable textual backdoor attacks via
  word substitution,''
\newblock in {\em {ACL/IJCNLP}}. 2021, pp. 4873--4883, Association for
  Computational Linguistics.

\bibitem{MSTBackdoor}
Fanchao Qi, Yangyi Chen, Xurui Zhang, Mukai Li, Zhiyuan Liu, and Maosong Sun,
\newblock ``Mind the style of text! adversarial and backdoor attacks based on
  text style transfer,''
\newblock in {\em EMNLP}. 2021, pp. 4569--4580, Association for Computational
  Linguistics.

\bibitem{liu2022piccolo}
Yingqi Liu, Guangyu Shen, Guanhong Tao, Shengwei An, Shiqing Ma, and Xiangyu
  Zhang,
\newblock ``Piccolo: Exposing complex backdoors in nlp transformer models,''
\newblock in {\em 2022 IEEE Symposium on Security and Privacy (SP)}. IEEE
  Computer Society, 2022, pp. 1561--1561.

\bibitem{TANN}
Yingqi Liu, Shiqing Ma, Yousra Aafer, Wen-Chuan Lee, Juan Zhai, Weihang Wang,
  and Xiangyu Zhang,
\newblock ``Trojaning attack on neural networks,''
\newblock in {\em 25th Annual Network and Distributed System Security
  Symposium, {NDSS} 2018, San Diego, California, USA, February 18-221, 2018}.
  2018, The Internet Society.

\bibitem{Zhai}
Tongqing Zhai, Yiming Li, Ziqi Zhang, Baoyuan Wu, Yong Jiang, and Shu{-}Tao
  Xia,
\newblock ``Backdoor attack against speaker verification,''
\newblock in {\em {ICASSP}}. 2021, pp. 2560--2564, {IEEE}.

\bibitem{canyouhear}
Stefanos Koffas, Jing Xu, Mauro Conti, and Stjepan Picek,
\newblock ``Can you hear it?: Backdoor attacks via ultrasonic triggers,''
\newblock in {\em Proceedings of the 2022 {ACM} Workshop on Wireless Security
  and Machine Learning}. 2022, pp. 57--62, {ACM}.

\bibitem{qiang2022opportunistic}
Qiang Liu, Tongqing Zhou, Zhiping Cai, and Yonghao Tang,
\newblock ``Opportunistic backdoor attacks: Exploring human-imperceptible
  vulnerabilities on speech recognition systems,''
\newblock in {\em Proceedings of the 30th ACM International Conference on
  Multimedia}. 2022, pp. 2390--2398, {ACM}.

\bibitem{shi2022audio}
Cong Shi, Tianfang Zhang, Zhuohang Li, Huy Phan, Tianming Zhao, Yan Wang, Jian
  Liu, Bo~Yuan, and Yingying Chen,
\newblock ``Audio-domain position-independent backdoor attack via unnoticeable
  triggers,''
\newblock in {\em Proceedings of the 28th Annual International Conference on
  Mobile Computing And Networking}. 2022, pp. 583--595, {ACM}.

\bibitem{Ye2019adversarial}
Yehao Kong and Jiliang Zhang,
\newblock ``Adversarial audio: A new information hiding method and backdoor for
  dnn-based speech recognition models,''
\newblock {\em arXiv preprint arXiv:1904.03829}, 2019.

\bibitem{nyquist1}
H.~Nyquist,
\newblock ``Certain factors affecting telegraph speed,''
\newblock {\em The Bell System Technical Journal}, vol. 3, no. 2, pp. 324--346,
  1924.

\bibitem{voiceprint1962}
Lawrence~George Kersta,
\newblock ``Voiceprint identification,''
\newblock {\em The Journal of the Acoustical Society of America}, vol. 34, no.
  5, pp. 725--725, 1962.

\bibitem{xvector}
David Snyder, Daniel Garcia-Romero, Gregory Sell, Daniel Povey, and Sanjeev
  Khudanpur,
\newblock ``X-vectors: Robust dnn embeddings for speaker recognition,''
\newblock in {\em 2018 IEEE international conference on acoustics, speech and
  signal processing (ICASSP)}. IEEE, 2018, pp. 5329--5333.

\bibitem{voiceconversion}
Seyed~Hamidreza Mohammadi and Alexander Kain,
\newblock ``An overview of voice conversion systems,''
\newblock {\em Speech Communication}, vol. 88, pp. 65--82, 2017.

\bibitem{li2021starganv2}
Yinghao~Aaron Li, Ali Zare, and Nima Mesgarani,
\newblock ``Starganv2-vc: {A} diverse, unsupervised, non-parallel framework for
  natural-sounding voice conversion,''
\newblock in {\em Interspeech 2021, 22nd Annual Conference of the International
  Speech Communication Association}. 2021, pp. 1349--1353, {ISCA}.

\bibitem{commonvoice}
Rosana Ardila, Megan Branson, Kelly Davis, Michael Kohler, Josh Meyer, Michael
  Henretty, Reuben Morais, Lindsay Saunders, Francis~M. Tyers, and Gregor
  Weber,
\newblock ``Common voice: {A} massively-multilingual speech corpus,''
\newblock in {\em {LREC}}. 2020, pp. 4218--4222, European Language Resources
  Association.

\bibitem{TSAA}
Steven~H Weinberger and Stephen~A Kunath,
\newblock ``The speech accent archive: towards a typology of english accents,''
\newblock {\em language and computers}, pp. 265--281(17), 2011.

\bibitem{deepspeech}
Awni Hannun, Carl Case, Jared Casper, Bryan Catanzaro, Greg Diamos, Erich
  Elsen, Ryan Prenger, Sanjeev Satheesh, Shubho Sengupta, Adam Coates, et~al.,
\newblock ``Deep speech: Scaling up end-to-end speech recognition,''
\newblock {\em arXiv preprint arXiv:1412.5567}, 2014.

\bibitem{deepspeech2}
Dario Amodei, Sundaram Ananthanarayanan, Rishita Anubhai, Jingliang Bai, Eric
  Battenberg, Carl Case, Jared Casper, Bryan Catanzaro, Qiang Cheng, Guoliang
  Chen, et~al.,
\newblock ``Deep speech 2: End-to-end speech recognition in english and
  mandarin,''
\newblock in {\em International conference on machine learning}. PMLR, 2016,
  pp. 173--182.

\bibitem{VCTK}
Christophe Veaux, Junichi Yamagishi, and Kirsten Macdonald,
\newblock ``Cstr vctk corpus: English multi-speaker corpus for cstr voice
  cloning toolkit,''
\newblock 2017.

\bibitem{speechcmd}
Pete Warden,
\newblock ``Speech commands: A public dataset for single-word speech
  recognition,''
\newblock {\em Dataset available from http://download. tensorflow.
  org/data/speech\_commands\_v0}, vol. 1, 2017.

\bibitem{LSTM}
Sepp Hochreiter and J{\"{u}}rgen Schmidhuber,
\newblock ``Long short-term memory,''
\newblock {\em Neural Comput.}, vol. 9, no. 8, pp. 1735--1780, 1997.

\bibitem{resnet}
Kaiming He, Xiangyu Zhang, Shaoqing Ren, and Jian Sun,
\newblock ``Deep residual learning for image recognition,''
\newblock in {\em {CVPR}}. 2016, pp. 770--778, {IEEE} Computer Society.

\bibitem{VGG}
Karen Simonyan and Andrew Zisserman,
\newblock ``Very deep convolutional networks for large-scale image
  recognition,''
\newblock in {\em 3rd International Conference on Learning Representations,
  {ICLR}}, 2015.

\bibitem{wideresnet}
Sergey Zagoruyko and Nikos Komodakis,
\newblock ``Wide residual networks,''
\newblock in {\em Proceedings of the British Machine Vision Conference}. 2016,
  {BMVA} Press.

\bibitem{srchallenge}
Elliott Julia, McDonald Mark, and Warden Pete,
\newblock ``Tensorflow speech recognition challenge,''
\newblock 2017.

\bibitem{crossentropy}
George Cybenko, Dianne~P O'Leary, and Jorma Rissanen,
\newblock {\em The Mathematics of Information Coding, Extraction and
  Distribution}, vol. 107,
\newblock Springer Science \& Business Media, 1998.

\bibitem{SGD}
Ning Qian,
\newblock ``On the momentum term in gradient descent learning algorithms,''
\newblock {\em Neural Networks}, vol. 12, no. 1, pp. 145--151, 1999.

\end{thebibliography}

\end{document}